\documentclass[twocolumn,preprintnumbers,amsmath,amssymb,a4paper,superscriptaddress,showpacs]{revtex4}
\usepackage{graphicx}
\usepackage{latexsym}
\usepackage{float}
\begin{document}

\title{Effect of disorder outside the CuO$_{2}$ planes on $T_{c}$ of copper oxide superconductors}

\author{K. Fujita}
\email{fujita@lyra.phys.s.u-tokyo.ac.jp}
\affiliation{Department of Advanced Materials Science, University of Tokyo, Tokyo 113-0033, Japan}

\author{T. Noda}
\affiliation{Department of Superconductivity, University of Tokyo, Tokyo 113-0033, Japan}

\author{K. M. Kojima}
\affiliation{Department of Physics, University of Tokyo, Tokyo 113-0033, Japan}

\author{H. Eisaki}
\affiliation{Nanoelectronics Research Institute, National Institute of Advanced Industrial Science and Technology (AIST), Ibaraki, 305-8568, Japan}

\author{S. Uchida}
\affiliation{Department of Advanced Material Science, University of Tokyo, Tokyo 113-0033, Japan}
\affiliation{Department of Superconductivity, University of Tokyo, Tokyo 113-0033, Japan}
\affiliation{Department of Physics, University of Tokyo, Tokyo 113-0033, Japan}

\date{\today}

\begin{abstract}
The effect of disorder on the superconducting transition temperature $T_{c}$ of cuprate superconductors is examined. Disorder is introduced into the cation sites in the plane adjacent to the CuO$_{2}$ planes of two single-layer systems, Bi$_{2.0}$Sr$_{1.6}$Ln$_{0.4}$CuO$_{6+\delta}$ and La$_{1.85-y}$Nd$_{y}$Sr$_{0.15}$CuO$_{4}$. Disorder is controlled by changing rare earth (Ln) ions with different ionic radius in the former, and by varying the Nd content in the latter with the doped carrier density kept constant. We show that this type of disorder works as weak scatterers in contrast to the in-plane disorder produced by Zn, but remarkably reduces $T_{c}$ suggesting novel effects of disorder on high-$T_{c}$ superconductivity. 
\end{abstract}


\maketitle

 High temperature superconductivity is realized in copper oxide materials upon doping charge carriers into the CuO$_{2}$ planes. Doping is made, in most cases, by chemical substitution or adding or removing oxygen atoms, which inevitably introduces disorder into the building blocks between the CuO$_{2}$ planes\cite{Eisaki}. Impact of such disorder on the electronic structure and the superconducting properties of high-$T_{c}$ cuprates is still poorly understood. Since $T_{c}$ is so strongly dependent on doping level, disorder appears to have a secondary effect. However, there are a few cases where it has or seems to have a dramatic effect. A familiar case is the partial Nd substitution for the La sites in La$_{2-x}$Sr$_{x}$CuO$_{4}$ (LSCO) which stabilizes the stripe order and suppresses superconductivity \cite{Crawford, Tranquada, Tajima}. Attfield \textit{et al}. also demonstrated that $T_{c}$ is systematically reduced with increasing cation disorder in the (La, Sr)O blocks in LSCO \cite{Attfield, McAllister}. They ascribe the $T_{c}$ reduction to hole trapping or to the tendency to stabilize a distorted structure, possibly related to the stripe order. In addition, the recent scanning tunneling spectroscopy (STS) studies of the surface electronic state of Bi$_{2}$Sr$_{2}$CaCu$_{2}$O$_{8+\delta}$ (Bi2212) has attracted much attention, and it is discussed that the disorder in the building blocks, BiO or SrO planes, might be a source of the observed electronic inhomogeneity\cite{McElroy1}.

A motivation of the present study is to quantitatively investigate the effect of disorder on $T_{c}$, specifically disorder in the building block next to the CuO$_{2}$ planes, using two single-layer systems, Bi$_{2}$Sr$_{2}$CuO$_{6+\delta}$ (Bi2201) and La$_{2-x}$Sr$_{x}$CuO$_{4}$. While the former does not show stripe instability, the latter is a prototypical one with competing stripe and superconducting order. The Bi-cuprates contain excess oxygen atoms in the BiO planes, and hole density in the CuO$_{2}$ plane is controlled by the amount of excess oxygen $\delta$. The doping is also controlled by partially replacing trivalent rare earth (Ln) for divalent Sr. It is known that Bi$_{2}$Sr$_{2}$CuO$_{6}$ with ideal cation compositioins is difficult to synthesize. In order to grow Bi2201 crystals, one needs to repalce Sr$^{2+}$ ions, Ln$^{3+}$ or Bi$^{3+}$ ions \cite{Ono,Banhofer,Nameki}, so the SrO plane is necessarily disordered in real crystals. The effect of disorder at Sr sites on the CuO$_{2}$ plane is expected to be stronger than the disorder in the distant BiO plane. In the  case of LSCO, when La is partially repalaced by Nd, in the form of La$_{2-x-y}$Nd$_{y}$Sr$_{x}$CuO$_{4}$ (LSCO), there is cation disorder due to the difference in the ionic radius among La$^{3+}$, Sr$^{2+}$, and Nd$^{3+}$.

\begin{figure}[tb]
\includegraphics[width=6cm]{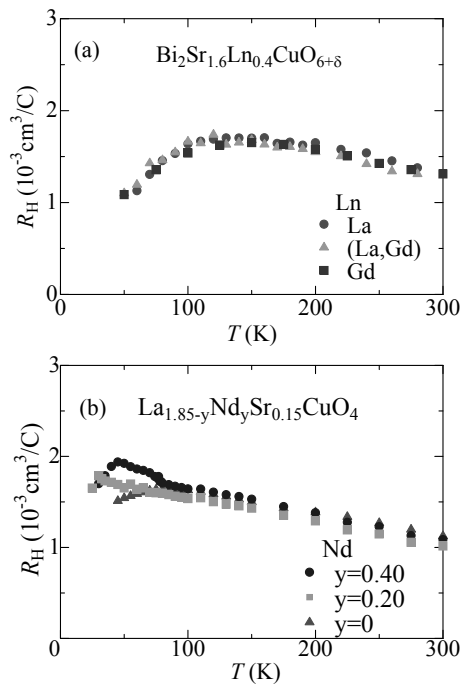}
\caption{Temperature dependece of the Hall coefficient for three samples of Ln-Bi2201 with different Ln (a), and for three samples of Nd-LSCO with $y=0$, $y=0.2$ and $y=0.4$(b). The data for other Ln's in Bi2201 and for different $y$'s in Nd-LSCO are basically the same as those shown here. }
\label{Hall}
\end{figure}

We have grown a series of Bi$_{2}$Sr$_{1.6}$Ln$_{0.4}$CuO$_{6+\delta}$ (Ln-Bi2201) and La$_{1.85-y}$Nd$_{y}$Sr$_{0.15}$CuO$_{4}$ (Nd-LSCO) single crystals. The degree of disorder is changed by varying Ln ions with different ionic radii in Bi2201, Ln=La, Nd, Eu, Gd, and in addition, we prepared Bi$_{2}$Sr$_{1.6}$La$_{0.2}$Gd$_{0.2}$CuO$_{6+\delta}$. For La$_{1.85-y}$Nd$_{y}$Sr$_{0.15}$CuO$_{4}$ disorder was varied by changing Nd$^{3+}$ content (y=0, 0.10, 0.12, 0.20, 0.40, and 0.60). All the samples in each series were prepared to have the same doping level by fixing the Ln content (=0.4) in Bi2201 and the Sr content (=0.15) in Nd-LSCO. In the case of Ln-Bi2201 the advantage of the Ln substitution is that it ensures Bi:(Sr+Ln):Cu=2:2:1 stoichiometry to within 1\%, which was checked by the inductively-coupled-plasma (ICP) spectroscopy. In fact, the $c$-axis lattice constant, a good measure of the stoichiometry and/or doping level, systematically decreases with increasing the (average) atomic number of Ln-from  24.44$\textrm{\AA}$ for La to 24.35$\textrm{\AA}$ for Gd. These values are in agreement with the previously reported values\cite{Nameki}. To assure uniform oxygen distribution and the same oxygen content, the sample were annealed under the same conditions, at 650$^{\circ}$C for 48h in flowing O$_{2}$ gas, and then quenched to room temperature. The excess oxygen content $\delta$ was estimated to be 0.3$\pm$0.1 by the iodometric titration,  but is difficult to accurately determine in the case of single crystalline Bi2201 by the usual method. The constant doping level or the constant carrier density was confirmed by measurement of the Hall coefficient (with current parallel to the planes and magnetic fields applied perpendicular to the planes), the result of which is shown in Fig.\ref{Hall} for three samples of Ln-Bi2201 and of Nd-LSCO\cite{Noda}. Both magnitude and temperature dependence of the Hall coefficient do not show significant change, indicating that the normal-state carrier density is almost the same.

\begin{figure}[tb]
\includegraphics[width=6cm]{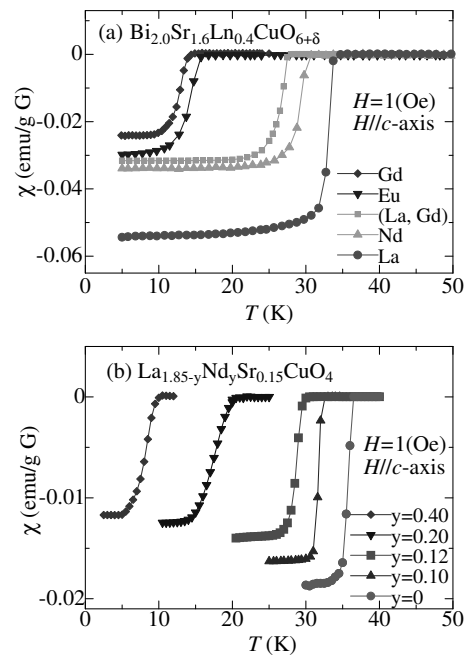}
\caption{Superconducting transitions measured by the field-cooled (Meissner) dc magnetic susceptibility for (a) Ln-Bi2201 and (b) Nd-LSCO.} 
\label{Tc}
\end{figure}

Figs.\ref{Tc}(a) and \ref{Tc}(b), show the field-cooled (Meissner) dc magnetic susceptibility for Ln-Bi2201 and Nd-LSCO, respectively. $T_{c}$ is found to appreciably decrease when La is substituted by other Ln in Bi2201 and when the Nd content increases in LSCO. Since the $T_{c}$ reduction is larger for less magnetic Eu than for magnetic Nd in Bi2201, the decrease is not due to magnetic pair breaking, but to disorder arising from difference in the Ln/Sr ionic radius. We see that the Meissner signal at low temperature also decreases with decreasing $T_{c}$.

\begin{figure}[tb]
\includegraphics[width=6cm]{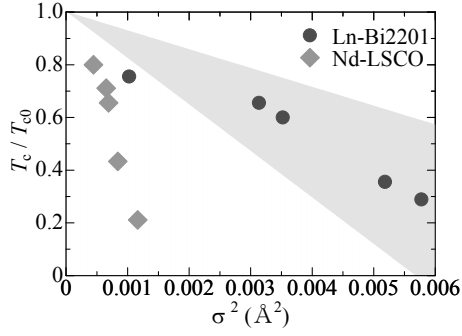}
\caption{Relation between normalized $T_{c}$ (by $T_{c0}$) and $\sigma ^2$ for the samples investigated in this work. $T_{c0}$ is estimated by extrapolating the data to $\sigma ^2 =0$ \cite{Shannon}. The previous results by Attfield $\textit{et al.}$\cite{Attfield, McAllister} are within the shaded zone.}
\label{Tc0}
\end{figure}

The ionic radius of Ln decreases monotonically with atomic number: from 1.216$\textrm{\AA}$ (La$^{3+}$) to 1.103$\textrm{\AA}$ (Gd$^{3+}$) which should be compared with the ionic radius 1.310$\textrm{\AA}$ of Sr$^{2+}$\cite{Shannon}, and thus one expects the SrO((La,Sr)O) planes in Bi2201(LSCO) to be more disordered with increasing Ln atomic number(the Nd content). Hence to make the analysis more quantitative, we define and evaluate the degree of disorder by the variance of ionic radius $r$ following Attfield \textit{et al.}  $\sigma ^2 \equiv$${\langle r^2 \rangle}-{\langle r \rangle}^2$, where $\langle \rangle$ denotes the site average, at the Sr and La sites for Ln-Bi2201 and Nd-LSCO, respectively. The degree of disorder is lowest for La-Bi2201 and Nd-LSCO ($y$=0), since the difference in the ionic radius between Sr$^{2+}$ and Ln$^{3+}$ is smallest for Ln=La, and actually the maximum $T_{c}$ values are realized for La-Bi2201 ($T_{c}$$\sim$ 35K) and Nd-free LSCO ($T_{c}$=39K). For Ln-Bi2201, $\sigma ^2$ increases in the order of La, Nd (La, Gd), Eu, and Gd ($\langle r \rangle$ decreases in the order La, (La, Gd), Nd, Eu, and Gd). For Nd-LSCO, $\sigma ^2$ increases with increasing the Nd content $y$. In Fig.\ref{Tc0} $T_{c}$ is plotted as a function of $\sigma ^2$ where $T_{c}$ is normalized by the value $T_{c0}$ ($\sim$45K) obtained by extrapolating the present results toward $\sigma ^2=0$\cite{Shannon}. The results of Attfield $\textit{et al.}$ \cite{Attfield, McAllister}, plotted in the same way, are within the shaded zone in Fig.\ref{Tc0}. It can be seen that the $T_{c}$ reduction rates are nearly the same, except for Nd-LSCO. The $T_{c}$ reduction rate in Nd-LSCO is significantly larger compared with other systems.

\begin{figure}[tb]
\includegraphics[width=6cm]{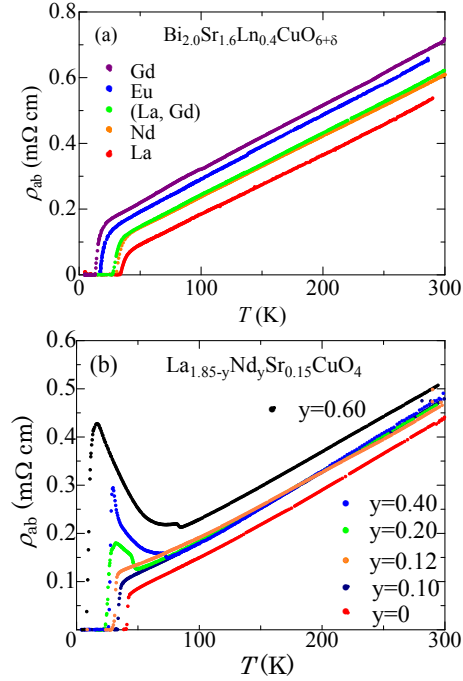}
\caption{(color). Temperature dependence of in-plane resistivity for optimally doped (a) Bi$_{2.0}$Sr$_{1.6}$Ln$_{0.4}$CuO$_{6+\delta}$ and (b) La$_{1.85-y}$Nd$_{y}$Sr$_{0.15}$CuO$_{4}$ in which disorder is introduced by changing the Ln species and the Nd content, respectively. }
\label{resistivity}
\end{figure}

Temperature dependences of the in-plane resistivity measured for the present samples are shown in Figs.\ref{resistivity}(a) and \ref{resistivity}(b). For Ln-Bi2201, all the resistivity curves show linear $T$ dependence from room temperature down to $T_{c}$, characteristic of high-$T_{c}$ cuprates near optimal doping. While the $T$-linear slope of resistivity does not change, giving additional evidence for the constant carrier density, the $T$-independent component $\rho _0$ (residual resistivity) increases with increasing $\sigma ^2$. This behavior is qualitatively similar to the Zn-doping effect on resistivity \cite{Fukuzumi}, but the magnitude of $\rho _0$ is much smaller. The present result is not in agreement with that reported by McAllister and Attfield \cite{McAllister}. They used polycrystalline samples to measure resistivity, but single crystals are crucial for quantitative analysis of resistivity in highly anisotropic materials. Together with the result of Hall effect, we conclude that no hole trapping takes place in the normal state by disorder, in contrast to the conclusion of McAllister and Attfield. The change of resistivity of Nd-LSCO (Fig.\ref{resistivity}(b)) is basically the same, except for the upturn at low temperatures observed for $y\agt$0.12 due to the development of stripe order.

\begin{figure}[tb]
\includegraphics[width=6cm]{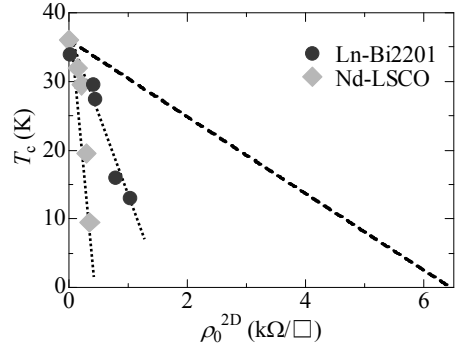}
\caption{$T_{c}$ vs disorder-induced ${\rho}_{0}^{2\rm{D}}$. The dotted curves connecting the data are guides for the eyes. For comparison the result for the Zn-doped LSCO with $x=0.15$ is shown by the dashed curve \cite{Fukuzumi}. }
\label{residual}
\end{figure}

Residual resistivity is a measure of the scattering rate of charge carriers by impurities or defects in a crystal. It is meaningful to make a comparison with the result for Zn doped cuprates. One finds that $\rho _0$ for Bi2201 with 20\% substitution of Gd for Sr (and $\rho _0$ for Nd-LSCO with 20\% Nd substituion ) is comparable with that for LSCO ($x=0.15$) with only 1\% Zn substitution for Cu\cite{Fukuzumi}. Therefore, the substituted Ln ions work as weak elastic scatterers in both Bi2201 and LSCO. On the other hand, the non-magnetic impurity Zn is a strong carrier-scatterer (almost in the unitarity limit) in the normal state, and only 3\% substitution for the in-plane Cu is enough to completely destroy superconductivity in La$_{1.85}$Sr$_{0.15}$CuO$_{4}$. In the case of Zn-doping, a universal relation between $T_{c}$ reduction and residual resistivity is found, and $T_{c}$ goes to 0 as Zn-induced ${\rho _{0}}^{\textrm{2D}}$ (two-dimensional residual resistivity per CuO$_2$ plane) approaches the value near the two-dimensional universal resistance $h/4e^{2}$ ($\sim$ 6.5 k$\Omega/\Box$) at which a superconductor-insulator transition takes place. Based on these observations, a phase-fluctution scenario was proposed by Emery and Kivelson to explain the $T_{c}$ reduction mechanism due to Zn \cite{EmeryKivelson}. To make a quantitative comparison, $T_{c}$ is plotted against ${\rho _{0}}^{\textrm{2D}}$ for the present systems in Fig.\ref{residual}, together with the data for the Zn-substituted La$_{1.85}$Sr$_{0.15}$CuO$_{4}$ shown by the dashed line \cite{Fukuzumi}. One can immediately see that the disorder outside the CuO$_{2}$ planes reduces $T_{c}$ as much as the Zn-doping does, even though it induces much smaller ${\rho _{0}}^{\textrm{2D}}$, suggesting that mechanism of the $T_{c}$ reduction is different from that due to Zn doping.

The low-temperature resistivity upturn and the preceding small jump seen for Nd-LSCO (Fig.\ref{resistivity}(b)) signals the formation of stripe order. B$\ddot{\rm{u}}$chner \textit{et al.} postulated that the stripes are pinned by the lattice distortions, specifically the tilt of the CuO$_{6}$ octahedra, when the tilt angle exceeds a critical value \cite{Buchner}. As the stripe order is competing with the superconducing order in this system, the $T_{c}$ suppression might be accelerated by this phase competition.

The $T_{c}$-suppression mechanism in Ln-Bi2201, and certainly in the material series studied by Attfield \textit{et al.}, would be different from that in Nd-LSCO. The difference might be linked to  the lattice distortions, possibly the tilting of CuO$_{6}$ octahedra, random local distortion in Ln-Bi2201, and long-range periodic distortion for Nd-LSCO, which would help to pin down the static stripe order. In the case of Ln-Bi2201, the substituted Ln does not induce a long-range periodic lattice distortion, instead the lattice distortion would be localized around each Ln site. From the relatively small residual resistivity resulting from this disorder, it is conceivable that the disorder may have a small effect on the carriers near the nodal points of Fermi surface (FS) which make a dominant contribution to the in-plane resistivity due to their large Fermi velocity and longer lifetime\cite{Zhou, Valla}. Scalapino and coworkers have recently analyzed the effect of impurities or defects outside the CuO$_{2}$ planes on the resistivity within the framework of Fermi liquid and BCS theory with a $d$-wave gap\cite{Scalapino}. The observed small residual resistivity is consistent with their analysis assuming elastic forward scattering of carriers from out-of-plane impurities. On the other hand, as we demonstrate in Figs.\ref{Tc} and \ref{resistivity}, the $T_{c}$ reduction is quite substantial due to this type of disorder. Such a $T_{c}$ reduction cannot be explained in the $d$-wave BCS framework with elastic forward scattering, and hence is suggestive of a novel effect of the disorder. 

One possibility is that it is related to the local lattice distortions. Local distortions, such as tilt of CuO$_{6}$ octahedra, would lead to local modulation of transfer integrals (overlap integrals of the electronic wavefunctions of neighboring atoms) in the CuO$_{2}$ plane. Experimentally, it becomes increasingly clear that the dominant contribution to the SC condensate comes from this antinodal region\cite{Feng,Ding}. Among the relevant transfer integrals, the nearest-neighbor hopping $t$, the next nearest one $t^{\prime}$, and the third nearest $t^{\prime \prime}$, $t^{\prime}$ most affects the electronic structure in the antinodal region. In momentun space, it can be easily illustrated that the FS topology is affected near the Brillouin zone edge (antinodal region) by a change of $t^{\prime}$, as the position of the flat band (or van Hove singularity) is very sensitive to $t^{\prime}$. Thus, we speculate that the disorder considered here locally modulates $t^{\prime}$. 

 We note that there are some reports that suggest a correlation between $T_{c}$ and $t^{\prime}$. Based on the LDA band calculation, Pavarini \textit{et al}. demonstrated that the ratio $t^{\prime}$/$t$ is determined by the axial orbital hybridization between Cu4$s$ and apical oxygen 2$p_z$\cite{Pavarini} (or between Cu 3$d_{3z^2-r^2}$ and apical oxygen 2$p_z$\cite{Raimondi}) and that the maximum $T_{c}$ of each material is correlated with the parameter $t^{\prime}$/$t$. The recent ARPES study has successfully estimated $t^{\prime}$ for Bi2212 and LSCO\cite{Tanaka} and found a correlation between $T_{c}$ and $t^{\prime}$ in agreement with the theoretical prediction. In the light of the recent STM/STS results\cite{McElroy}, it is also possible to speculate that local modulation of $t^{\prime}$, that is, local depression of the flat band near ($\pi ,0$)/($0,\pi$) below $E_{\textrm{F}}$, makes nanoscale non-SC regions in the CuO$_{2}$ plane, leading to an overall decrease in the superfluid density $\rho _s$ and hence to reduction in $T_{c}$ via the Uemura relation\cite{Uemura}. In fact, the decrease of the Meissner signal associated with the reduction of $T_{c}$ (Fig.\ref{Tc}) is suggestive of a decrease in $\rho _s$ \cite{muSR}. One needs more systematic and combined experiments of ARPES, STM/STS, and $\mu$SR, to obtain evidence for the changes of local $t^{\prime}$ and global $\rho _s$ by disorder investigated here. 

The present results show the presence of a novel effect of disorder residing outside the CuO$_{2}$ planes on high-$T_{c}$ superconductivity which is distinct from the so-far-known effects, such as change of doping level and Zn doping. The results indicate that the disorder in the building blocks play a more important role than previously anticipated in high-$T_{c}$ cuprates, and thus point towards a direction of improving $T_{c}$ by controlling disorder there.

We thank A. Fujimori, M. Ogata, E. Pavarini, R. B. Laughlin, T. Tohyama, J. C. Davis, and Z. -X. Shen for useful discussions. This work was supported by the 21st Century COE program from the Japan Society for the Promotion of Science (JSPS).


\begin{thebibliography}{100}
\bibitem{muSR}We have performed the experiment on La-Bi2201 ($T_{c}$=34K) and Eu-Bi2201 ($T_{c}$=16K). Preliminary result indicates that $\rho _s$ for Eu-Bi2201 is nearly half of that for La-Bi2201.
\end{thebibliography}
\end{document}